# Influence of magnetic field on evaporation of a ferrofluid droplet

Mudra Jadav, (late) R.J.Patel and R.V.Mehta*

Influence of magnitude and direction of static magnetic field applied on a drying drop of a laboratory synthesized water base ferrofluid placed on a plane glass plate is investigated. Like all other multi-component fluid this drop also exhibited familiar coffee ring pattern in absence of field while, this pattern is modulated by applying magnetic field and thickness of the ring modulated by the applied field and thickness of the ring decreases exponentially with increase in field when direction of field is normal to the plane of the substrate. Effect of the field on the evaporation rate and temporal variation of contact angle is also studied. Results are analysed in light of available models. The findings may be useful in applications like ink-jet printing, painting and display devices involving ferrofluids.

## Introduction

It is common experience that a drop of tea or coffee left a ring pattern on the edge of the drop when allowed to evaporate. This pattern called coffee ring is ubiquitous in nature and observed in almost all types of fluids ranging from liquids like wine, salt solution, blood, nanofluids and coarse dispersions. This effect is undesirable for several applications like printing, painting, thin film deposition, biological labelling and like where uniform deposition is required. Hence, understanding of mechanisms of evaporation in drying drops is both scientifically interesting and technologically useful. First systematic study in this direction was carried out by Deegan and his co-workers [1-3] .They ascribe the physical origin of the coffee ring pattern due to a combination of pinning of the three-phase contact line and a radial capillary flow from centre to the edge in which dispersed particles are also drag alongwith carrier liquid. This theory was later modified by incorporating shape, concentration of dispersed phase, thermal effect etc. [4-6]. Yunker et al have shown that when a drop of a suspension of ellipsoidal particles or a suspension of spherical particles containing a small amount of ellipsoidal particles of same composition is dried in ambient conditions then coffee ring effect

___________________________________________

Department of Physics, Maharaja Krishnakumarsinhji Bhavnagar
University, Bhavnagar, 364002 Gujarat, India..
rvm@mkbhavuni.edu.in

effect can be suppressed and more uniform distribution of particles can be obtained. This is attributed to the strong interparticle capillary interactions caused by deformation of the interface due to anisotropic shape of the ellipsoidal particles [6]. Applications of alternating electric field as well as subjecting the colloid to acoustic field are also used to suppress the ring structure [7, 8]. Over and above these, there are several other methods like pH control, contact angle hysteresis of substrate, addition of surfactant, drop size control etc. which are reported to be useful to reverse the ring structure [9-12]. Recently charged particles are used to control ring pattern [13, 14]. Several theoretical papers and computational analysis also published [15-17]. It may be noted that technique which is suitable for one type of fluid may not be appropriate for other fluids. Coffee ring patterns id Dispersions which contain magnetic or magnetizable particles can be suppressed by applying magnetic force. This was demonstrated in water based suspensions of monodispersed micron sized ellipsoidal particles of hematite crystals coated with SDS. It was shown that coffee ring effect can be manipulated by applying static as well as alternating magnetic fields [18]. Suspension of Janus magnetic particles is another example in which magnetic field is used to modulate ring patterns. A Janus particle is composite of two materials say silver and iron oxide. Physical properties of both are different. Such particles have also varied applications. Mathematical modelling and computational analysis of evaporating drop of Janus particles is studied by several workers [19-21]. Amongst all such magnetic dispersions magnetic fluid also known as ferrofluid has well established branch of science and technology.

Ferrofluid is composed of either sterically or charged stabilized nanomagnetic particles dispersed in a carrier liquid like water, hydrocarbons etc. At nano size magnetic particle has only one domain hence it acts as a nanomagnet. Under normal conditions colloidal dispersion of these particles Brownian motion prevails and it behaves as normal fluid. Under action of applied field the fluid performs like superparamagnetic fluid because of large magnetic moment compared to paramagnetic molecules and follows Langevin theory. Science and technology of the fluid is widely exploited [22-24] .Extensive work on applications of this fluid in areas of biotechnology, biomedicine, nanotechnology nanophotonics etc. are being still carried out [25-27]. In applications like ink-jet printers, ferrofluid thin films, biosensors and like where only a drop of ferrofluid is used, understanding of coffee ring effect and its control under application of applied field is necessary [28-30]. Only a few papers are available on detailed experimental investigations on this aspect [28]. In the present paper we study effect of externally applied uniform magnetic field on distribution of particles when a drop of water base ferrofluid is evaporated at room temperature. Under zero fields usual coffee ring pattern is observed. This pattern is modulated when a field is applied. When the field is normal to the surface of the substrate the pattern is suppressed at a critical value of the field. Effect of the field on the evaporation rate and contact angle is also studied.

**Materials and Methods**

Magnetic particles were prepared by co-precipitating divalent and trivalent iron ions by alkaline solution and treating under hydrothermal conditions. $FeSO_4.7H_2O$, 27 g, in 100 ml doubly-distilled water and 57 g $FeCl_3.6H_2O$ also in 100 ml water were thoroughly mixed and added to 8 M $NH_4OH$ under continuous stirring at room temperature. The precipitates were heated at 80°C for 30 min. The pH was maintained approximately at 10 by the addition of $NH_4OH$ during the reaction. Particles thus obtained were black and exhibited a strong magnetic response. Impurity ions such as chlorides and sulphates were removed by washing with copious amounts of hot distilled water. Water based stable dispersion of the above particles was obtained by bilayer coating of oleic acid on the particles [Figure 1]. If a monolayer of oleic acid is coated on the magnetite particles then surface becomes hydrophobic and dispersion will not be stable in water. To convert these to hydrophilic one, second layer of the same surfactant is used [31]. Aggregates if any were removed by centrifuging the above fluid at 8000 RPM for 20 minutes. Magnetization measurement for this stable magnetic fluid was carried out using search coil method. Saturation magnetization, particle diameter and volume fraction of the fluid were found to be 130 G, 12nm and 0.091 respectively. Dynamic light scattering (DLS) measurements were also taken for the fluid. Using DLS, Particle diameter and polydispersity factor were found to be 12nm and 0.42 respectively. For the experimental measurements, the fluid was diluted with different amounts of double distilled water. It was confirmed that the fluid was not dilution sensitive.

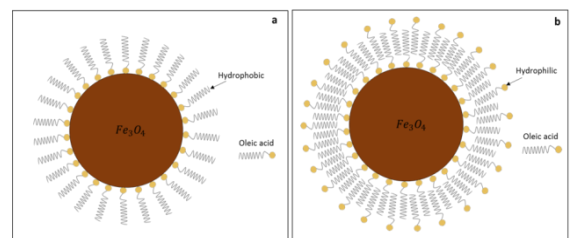

Fig. 1 Schematic of method used to convert hydrophobic magnetite particle to hydrophilic by bilayer coating of oleic acid.

A micropipette having capacity to control volume from 0.2µl to 50µl was used to place small drops of ferrofluid on a clean microscopic glass slide. It may be remarked here that if the glass substrate is treated with a surfactant like SDS then surfactant- induced Marangoni flow can also modulate the coffee ring pattern of the colloidal particles [6]. In the present work substrate was not treated with any such liquid and drop was directly deposited on it. The drop was allowed to evaporate under ambient conditions. After some time, the drop was fully dried and was examined under optical microscope (Magnus MLX). The microscope was attached with a CCD camera (YOKO) having 640 × 480 pixels resolution. Images were captured using 4× magnification lens. The field of view in the microscope using this lens was 0.71 mm. Calibration of the lens and Image analysis were carried out using Image-J open source software. Quantitative measurements were carried out using video microscopy with HANDYCAM 5 MP camera.

Experiments were carried out using drop sizes varying from 0.2 µl to 30µl and concentration varying from 2% to 25% of the aqueous ferrofluid. For each drop the experiment was repeated for three different conditions i.e. no applied magnetic field (H=0), magnetic field applied parallel to the substrate ($H_{||}$) and magnetic field applied perpendicular to the substrate ($H_\perp$).

Magnetic field in direction parallel to the plane of the substrate was applied by placing the substrate at the centre of pair of Helmholtz coils and field was varied by varying current from a constant current power supply. For perpendicular direction rare earth magnet was placed at the bottom of the substrate and distance was varied to vary the field. In both the case uniformity of the field and magnitude of the field strength was measured by a Gaussmeter having accuracy of ±1 Gauss. Field strength was varied from 0.05 T to 0.15 T. Weight loss was measured by using the microbalance (Shimadzu Corp. Model AEU-210) having accuracy of ±0.1mg. Ferrofluid drop placed on a microscope cover slip (thickness < 0.5mm) was mounted directly on the balance. For measurements with magnetic field we used a small Halbach magnet having magnetic field ~ 0.11 T on a surface while other surface was having almost zero field. Ferrofluid drop placed on a microscope cover slip (thickness < 05mm) was centrally arranged on this magnet. To increase distance between the magnet and balance (so as to minimize effect of the field on the balance if any) a nonmagnetic spacer was kept on the balance and on this the magnet alongwith the ferrofluid drop was placed. Weight loss during the evaporation was directly recorded.

**RESULTS**

Figure 2(a) shows CCD image of the dried drop having 0.2 µL volumes and 5% concentration and dried under zero field. Ring pattern is clearly observed indicating that Marangoni flow is negligible and situation satisfy all the conditions first listed by Deegan et al, *viz (1) the solvent meets the surface at a non-zero contact angle, (2) the contact line is pinned to its initial position as is commonly the case, and (3) the solvent evaporates* [ 3 ]. It was also observed that the ring thickness varies linearly with density of the fluid [not shown here]. Effects of applied magnetic field (0.15 T) on the pattern are shown in figure 2(b, c). When the field was parallel to the plane of the substrate ring pattern is replaced by long chains parallel to the plane of the substrate. Chains are more sharp on the right hand side of the periphery (b) .This may be due to either slight Inhomogeneities of the field or due to levelling of the substrate. When the field is applied in perpendicular direction again ring disappears but now smaller chains are formed with orientation appears to be normal to the substrate and their distribution is spread throughout the drop area. (c).

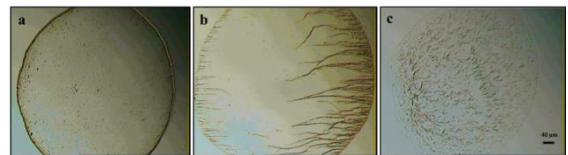

Fig.2 Effect of applied magnetic field on coffee ring pattern (a) H=0, (b) $H_{||}$= 0.15T and (c) = 0.15T. Drop volume = 0.2µl.

Figure 3 (a-f) shows effect of variation of field for parallel configuration. For better resolution only the area near the boundary is shown. At lower field strength (i.e.0.005T and 0.02T), some of the particles deposit at the drop edges to form thick ring pattern and some are combined into chains. But at higher field strength (i.e.0.07T, 0.10T) thickness decreases drastically and chains moves towards the central area. At the highest field (0.15 T) almost no deposition is observed at the edge. The long chains are not capable enough to travel to the edge and concentrate at the centre [Figure 4].Above experiments were repeated for perpendicular field configuration (Figure 5a - 5f). It is observed that, from 0.005T to 0.07 T, tendency of the particles to deposit at the boundary decreases. It becomes less dense as field reaches 0.10 T and small buckled chains deposit almost uniformly throughout the drop area. At 0.15 T almost no particle deposition at the periphery is observed.

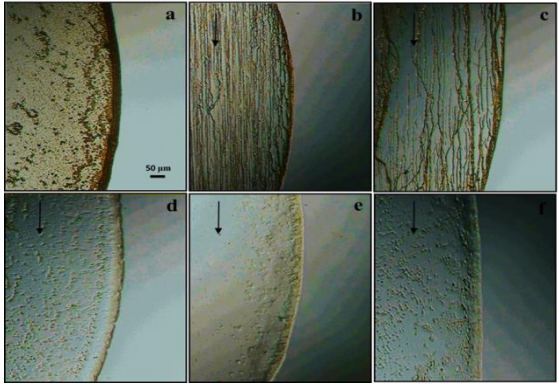

Fig.3 Effect of Variation of applied field $H_{||}$ on the ring thickness (a) 0.0 0 T (b) 0.005T (c) 0.02T (d)) 0.07T (e) 0.1T (f) 0.15T. Drop volume = 0.5μl and concentration= 3%

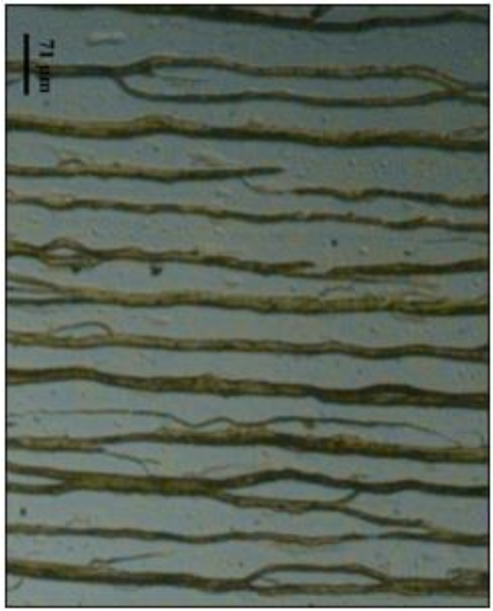

Fig.4 Microscopic image of thick chains observed at the centre of the dried drop when H||= 0.15T

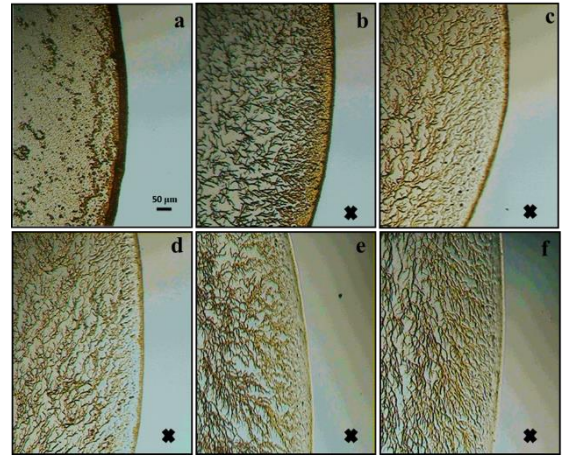

Fig: 5 Effect of variation of applied field $H_{\perp}$ on the ring thickness (a) 0 T (b) 0.005T (c) 0.02T (d) 0.07T (e) 0.1T (f) 0.15T. Drop volume = 0.5μl and concentration= 3%

Variation of the ring thickness with field strength for perpendicular configuration is plotted in figure 6. It is observed that the thickness decreases exponentially with increase in field strength.

Evaporation rate and contact angle variations are also important parameters in investigations of drying drops. We have also studied effect of applied field on these parameters. Figure 7 compares the weight loss (which is assumed to be proportional to the evaporation rate) of two identical drops measured at different times, with or without field.

To investigate effect of field on contact angle we have again take two identical drops and taken video graph of the images for zero field and field of 0.15 T. Image analysis was carried out to calculate contact angle at different times. Figure 8 shows the result. Temporal variations of the contact angle appear to be almost independent of the applied field. The results are discussed in the following section.

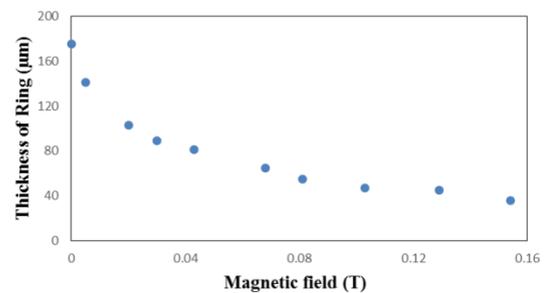

Fig.6 Plot of variation of thickness of the rings with increase in field strength. Field was perpendicular to plane of the substrate.

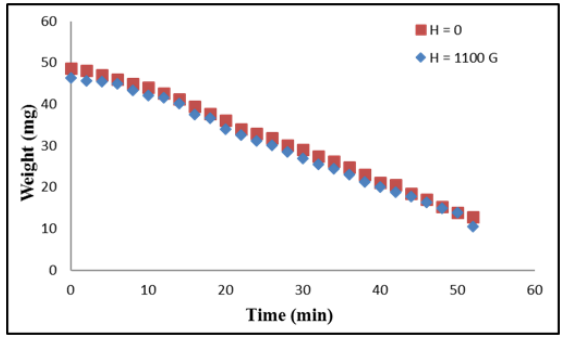

Fig.7 Plots of weight loss vs. Time when ferrofluid drops are evaporated with field H$_|$=0.11T and H=0T.

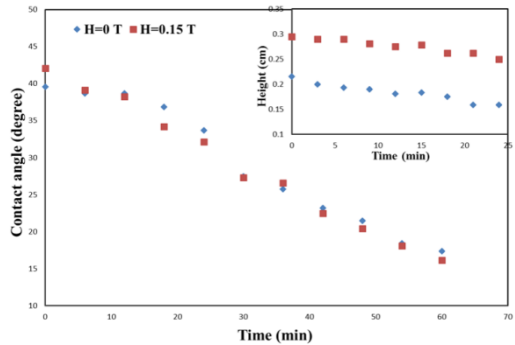

Fig.8 Variation of receding contact angle of the evaporating drops with time with and without field. Inset variation of drops height with time.

**Discussion**

A drop of colloid, placed on a plane solid surface, if allowed to evaporate under ambient conditions then in almost all cases it is observed that dispersed phase forms ring shape pattern at the edge of the drop. This coffee ring pattern arises due to pinning of contact line on the surface and resulting convective flow of the colloid towards the contact line. Ultimately liquid phase will evaporate and solid phase will be deposited along the contact line. This can be prevented if either contact line is not allowed to pin or dispersed phase is not allowed flow towards the contact line. As described earlier there are several methods to prevent formation of the ring patterns. None of these are suitable for a pure ferrofluid (without adding additional surfactant or ellipsoidal particles) in which sterically stabilized spherical magnetite are dispersed in water. It is well known that a ferrofluid will perform like an ordinary fluid in absence of an external field. Its physical properties will drastically change in presence of a magnetic field of moderate magnitude [22]. It is also shown that the properties depend upon whether the field is static, dynamic, uniform or inhomogeneous. They also depend on the direction of applied field with reference to the surface of the fluid.

The present experimental findings are interpreted on the basis of above work. Under absence of the field the ferrofluid drop behaves like an ordinary fluid and exhibits ring pattern upon evaporation of water (Figure 2a). It is also observed that over and above a densely packed dark ring at the contact line there are a number of randomly orientated chains and aggregates in the field of view. Random chains in absence of field were first predicted by de Gans and Pincus and later analysed both theoretically and experimentally by several workers [32-36]. The chains formed under zero fields were attributed to the dipole-dipole interaction between the particles which compete with the thermal motion of the particles. Further it was shown that when the fluid is subjected to an external magnetic field this chain formation will be enhanced and longer chains parallel to the direction of the field will be formed [37-39]. Microphotographs shown in figures 2- 5 also show random chains in zero fields along with random chains and enhancement of chain formation under the influence of the field. Additional feature observed is effect of field magnitude and direction on the particle density at the contact line. In both the directions of field it decreases with field and ultimately it reaches to minimum (Fig.3f). Thick and long chains are observed near the centre (Fig.4). It has been shown that if Marangoni flow is dominated that radial flow from the edge of the drop towards centre will increases [6a, b]. Marangoni flow is also dependent on the concentration of solid phase at the surface. On line of above theories we attribute above results to magnetic field induced Marangoni flow. When applied field is perpendicular to the substrate then also chain form parallel to the direction of the field and perpendicular to the substrate. These chains motion is blocked by the field and their distribution along the surface will be more uniform and ring formation will be blocked (Figure 5). In figure 6 variation of thickness of ring with field is plotted. The thickness exponentially decreases with field.

In applications like ink-jet printing, painting or fuel additive evaporation rate of a drying drop of ferrofluid is an important parameter. It can be determined by measuring weight loss at different time intervals [40]. When alternating magnetic field is applied magnetic particles in a ferrofluid can be heated by magneto relaxation heating [41]. It was also shown that vaporization rate is higher when alternating field is applied [41]. In the present case we have applied uniform magnetic field hence rotation of particles which is largely responsible to produce thermal diffusion in the fluid does not occur. Consequently rate of weight loss and hence evaporation rate does not change significantly with application of the field (Figure 7).

Deegan and co-workers have explained that coffee ring pattern arises due to contact angle hysteresis and capillary flow [2]. As shown above, in case of ferrofluid drop this effect can be suppressed and uniform distribution of particles can be achieved when applied field is perpendicular to the plane of the substrate. Coffee ring hysteresis can be measured by different methods [42]. One of these is evaporation method where receding contact angle is measured while drop is evaporating [43]. Using this method we have measured drop height as well as calculated receding contact angles at different time intervals. It is observed that i) For $H_\perp$ =0.15T the drop height is larger than that at H=0, ii) the receding contact angle at $H_\perp$ =0.15T is greater than that at H=0 and iii) rate of decrease of the contact angles is larger than that for zero field.[Fig. 8]. It is known that when a magnetic force is applied from the bottom of a ferrofluid drop it pushes the drop upwards and height of the drop increases. It is also reported earlier that contact angle also increases when field is applied [43, 44]. After 50 minutes carrier liquid (water) is completely evaporated and only solid phase remains. So now contact angle is only between solid particles and solid substrate. It is interesting to note that now the receding contact angle with field is slightly less than that without the field. Qualitatively this can be explained as follows. At the contact line there are three phases solid particles, carrier liquid and surrounding air (gas). It is shown that these solid particles also contribute to the pinning of contact line [43]. Evaporation rate is also larger at the contact line than at the centre. In case of ferrofluid drop too when field is not applied pinning of contact angle will take place and during evaporation the dispersed particles will also flow towards the pinned contact line. During the evaporation drop height continuously deceases and contact angle too also will decrease. When magnetic field is applied magnetic force will counteract the capillary flow and less number of particles will move towards the rim. Here too, height of the drop will decreases as water evaporates. This yield lower contact angle when field is applied. Quantitative analysis of these aspects and further experimental work are being carried out and will be published in next paper.

CONCLUSIONS:

We have studied, evaporation of a ferrofluid drop placed on a horizontal glass plate and subjected to a uniform magnetic field. It is demonstrated that structure and distribution of nanomagnetic particles in the dried drop depends on direction and magnitude of applied field. It is shown that thickness of pattern formed at the edge of the drop-also called coffee ring decreases exponentially with field when direction of the field is perpendicular to the direction of plane of the glass plate. Evaporation rate does not depend much on application of uniform magnetic field while the receding contact angle under the influence of magnetic field is less than the angle measured in absence of the field.

ACKNOWLEDGEMENTS:

Authors thank (Late) Dr. Rajesh Patel who has initiated this work. MJ is thankful to Miss Hiral Virpura for her help in synthesis of the ferrofluid. We also thank Professor S.P. Bhatnagar for helpful discussion. One of us (MJ) is thankful to Department of Science and Technology Government of India, New Delhi for INSPIRE fellowship.